\documentclass[sigconf,nonacm]{acmart}
\setcopyright{none}   
\makeatletter
\renewcommand\@copyrightpermission{} 
\makeatother
\AtBeginDocument{%
  }

\copyrightyear{2025}
\acmYear{2025}
\setcopyright{rightsretained}
\acmConference[ASPDAC '25]{30th Asia and South Pacific Design Automation Conference}{January 20--23, 2025}{Tokyo, Japan}
\acmBooktitle{30th Asia and South Pacific Design Automation Conference (ASPDAC '25), January 20--23, 2025, Tokyo, Japan}\acmDOI{10.1145/3658617.3697777}
\acmISBN{979-8-4007-0635-6/25/01}

\usepackage{subcaption}
\usepackage{amsmath,amsfonts}
\usepackage{algorithmic}
\usepackage{algorithm}
\usepackage{comment}
\usepackage{xspace}
\usepackage{enumitem}

\begin{document}


\title{TimingLLM: A Two-Stage Retrieval-Augmented Framework for Pre-Synthesis Timing Prediction from Verilog}


\author{Armin Abdollahi}
\email{arminabd@usc.edu}
\orcid{0009-0007-1387-0995}
\affiliation{%
  \institution{University of Southern California}
  \city{Los Angeles}
  \state{CA}
  \country{USA}
}

\author{Negin Ashrafi}
\email{ashrafin@usc.edu}
\orcid{0009-0003-8414-2996}
\affiliation{%
  \institution{University of Southern California}
  \city{Los Angeles}
  \state{CA}
  \country{USA}
}

\author{Mehdi Kamal}
\email{mehdi.kamal@usc.edu}
\orcid{0000-0001-7098-6440}
\affiliation{%
  \institution{University of Southern California}
  \city{Los Angeles}
  \state{CA}
  \country{USA}
}

\author{Massoud Pedram}
\email{pedram@usc.edu}
\orcid{0000-0002-2677-7307}
\affiliation{%
  \institution{University of Southern California}
  \city{Los Angeles}
  \state{CA}
  \country{USA}
}

\renewcommand{\shortauthors}{Abdollahi et al.}

\begin{abstract}
\vspace{-2pt}
Early, tool-free prediction of post-synthesis timing remains a key obstacle to rapid RTL iteration. We introduce TimingLLM, a two-stage retrieval-augmented LLM pipeline that estimates worst negative slack (WNS) and total negative slack (TNS) directly from Verilog. Stage~1 is a fine-tuned LLM that acts as a compact post-synthesis timing oracle, producing path-level arrivals/required times that are summarized into lightweight structural–timing cues (e.g., bag-of-gates counts, critical-path depth, gate-type patterns). Stage~2 is an LLM-based regressor that predicts WNS/TNS and applies a learned diagonal steering vector at the last transformer block, computed from the $k$ nearest timing-labeled modules in a disjoint retrieval bank. On VerilogEval, TimingLLM attains $\mathrm{R}_{\text{WNS}} = 0.91$ (MAPE 12\%) and $\mathrm{R}_{\text{TNS}}=0.97$ (MAPE 16\%) while running 1.3–1.6$\times$ faster than prior methods. Training uses a new 60k-module Verilog corpus with synthesis reports, which we will release. After training once, TimingLLM can be adapted to new technology libraries and PVT corners by refitting only a small regression head on 1000 labeled modules per setting, consistently outperforming state-of-the-art baselines.
\end{abstract}




\maketitle

\section{Introduction}

The register–transfer level (RTL) is the earliest point in the design flow where functionality is fixed yet structural freedom remains high \cite{ref5, ref1}. Decisions made here dominate the eventual power, performance and area (PPA) of an application-specific integrated circuit \cite{ref6}. Accurate static timing analysis (STA) is unavailable at RTL because commercial tools require a post-synthesis netlist and detailed parasitics, so designers are forced to iterate blindly or wait hours for a full synthesis and layout run before discovering that their module violates the clock period \cite{ref7, ref8, ref17}. This lack of early feedback slows architectural exploration and compromises time-to-market \cite{ref9, ref10, zhao2025codev}. Recent large language model (LLM) \cite{ref1, ref2, ref3, ref4, abdollahi2026hdlforge} research has shown impressive ability to parse and generate synthesizable Verilog, yet these efforts focus almost exclusively on functional correctness and ignore timing closure. More broadly, data-driven modeling has shown promise across several application domains \cite{sun2025optimizing, jin2025novel}, motivating the use of predictors for complex engineering tasks such as early RTL timing estimation.

Parallel work on presynthesis PPA modelling tackles the timing
problem by converting RTL into graph or operator representations
and applying graph neural networks (GNNs) or analytical models \cite{ref12}. State-of-the-art systems such as MasterRTL and RTL-Timer achieve respectable correlation on worst negative slack (WNS) and total negative slack (TNS) but require heavy feature engineering, extensive retraining across technology libraries, or only deliver coarse module-level estimates \cite{ref23, fang2024annotating}. Moreover, these treat each design in isolation and therefore fail to exploit the growing repositories of RTL whose timing labels are already known \cite{ref8}.

This paper couples an LLM which reads RTL and sketches STA-style path-level timing with a lightweight predictor of the global WNS/TNS, aiming to deliver fast, early RTL estimates that closely track post-synthesis STA and guide design iteration. Final timing closure and sign-off continue to rely on full STA. In this paper, we propose an LLM-based WNS and TNS predictor. Stage~1, which is a fine-tuned foundational LLM model responsible for generating a compact STA-style timing report directly from RTL. This report, then, is converted into an $\ell_{2}$-normalized fingerprint that summarizes structural and timing cues such as gate and flip-flop counts, gate-type histograms, critical-path depth and endpoints which will be used only to index a disjoint retrieval bank and select the top-$k$ neighbors. Stage~2 is an LLM-based regressor whose architecture contains a fine-tuned LLM model with a small MLP head to map last-layer features (logits) to $(\mathrm{WNS},\mathrm{TNS})$.
The neighboring (retrieved) Verilog modules are individually processed as inputs to the LLM, and their logits are used to steer the logits of the target Verilog module being analyzed. The steered logits are then fed into an MLP to estimate the timing delay parameters.
This approach enables the neighbors to softly bias the query representation, increasing its similarity to representations observed during training, thereby improving prediction accuracy. Contributions of this paper can be summarized as follows:


\begin{enumerate}
  \item We introduce the first tool-free LLM framework that
        couples path-level STA reasoning with retrieval-guided
        regression to predict both WNS and TNS at RTL.
  \item To our knowledge, TimingLLM is the first use of a retrieval-conditioned diagonal steering residual injected at the last transformer block for RTL timing prediction, letting nearest neighbors guide internal features rather than being appended as tokens, which reduces noise and preserves the query’s syntax.
  \item On the VerilogEval benchmark, the proposed
        method boosts WNS correlation by up to 7 percentage
        points and cuts mean absolute percentage error by up to
        10 points over SOTA, all with small runtime.
  \item We curate and release a diverse dataset of 60k RTL modules with ground‑truth WNS and TNS labels, covering simple to complex designs, filtered and cleaned across various module types, to enable future RTL timing research and benchmarking.  

  \item we show that a single trained TimingLLM can be adapted to new libraries and PVT corners by refitting only a small regression head on 1000 labeled modules per setting, without retraining the timing reasoner or rebuilding the retrieval bank.
\end{enumerate}

\vspace{-0.3cm}
\section{Related Work}

Prior work has explored data-driven optimization and hardware-efficient architectures \cite{abdollahi2025icd} and accelerators for computation \cite{abdollahi2024menage}. These efforts highlight the potential of combining learning-based methods with domain-specific reasoning, which we extend to RTL-level timing prediction. Static timing analysis becomes fully accurate only after synthesis and placement, yet architectural choices at the register transfer level strongly influence PPA. Pre synthesis estimators translate HDL to graphs or operator representations and train graph neural networks. RTL Timer annotates slacks on Verilog with a timing engine heuristic plus a graph regressor \cite{fang2024annotating}. MasterRTL augments operator graphs to improve cross technology transferability \cite{ref23}. These methods require heavy feature engineering and repeated retraining across libraries and they treat each design in isolation. LLSM moves toward metric prediction by fusing AIG features with an LLM text embedding but trains a single global regressor and relies on explicit synthesis during data collection \cite{huang2025llsm}. CircuitFusion fuses HDL structural graphs and LLM generated functionality summaries in a multimodal encoder \cite{ref202}. DeepCircuitX aggregates thousands of open source projects with post synthesis PPA labels and shows that chain of thought supervision improves metric reasoning \cite{li2025deepcircuitx}. RocketPPA uses an LLM to predict power, delay, and area directly from HDL and, when trained on tens of thousands of designs, demonstrates that larger training sets improve PPA prediction on unseen Verilog modules \cite{abdollahi2025rocketppa}. Our method uses a two-stage LLM pipeline combining STA-style reasoning with retrieval-guided regression.

\section{Proposed LLM-based PPA-Estimator Method}
\subsection{Dataset generation}
To develop an efficient LLM-based framework, a comprehensive dataset is required. In this paper, we utilize the PyraNet~\cite{ref203} dataset, which contains both synthesizable and non-synthesizable Verilog modules. Since TimingLLM requires synthesizable Verilog modules, we implement a selection process to identify suitable modules and construct our dataset.
It should be noted that TimingLLM can also be applied to non-synthesizable Verilog code. However, since there are no reference timing reports available for comparison, we cannot evaluate its effectiveness on this type of code.

\begin{figure}[h]
    \centering
    \includegraphics[width=0.70\linewidth]{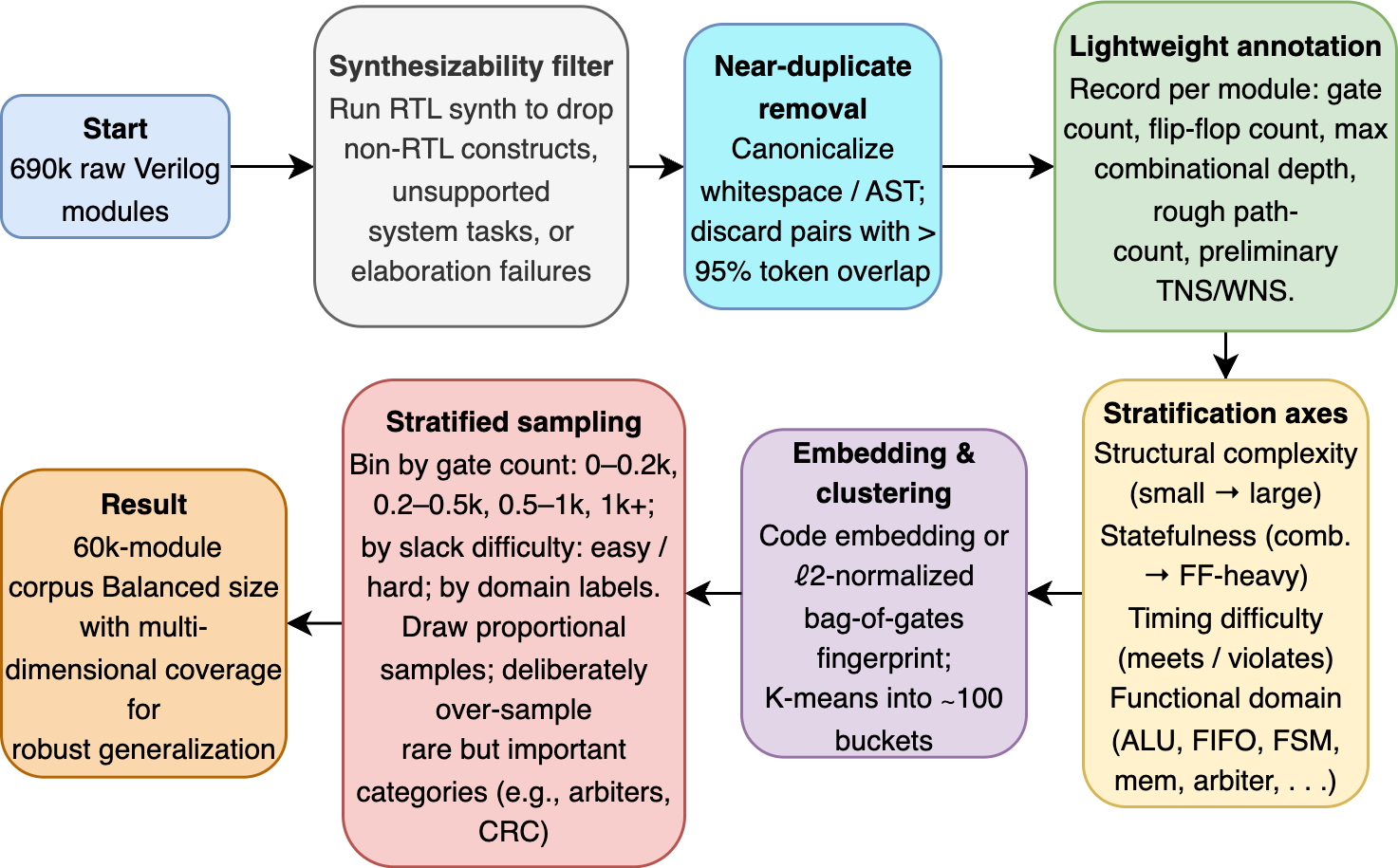}
    \captionsetup{justification=centering}
    \caption{Dataset curation and stratified selection pipeline for the 60k-module corpus.}
    \label{fig:3a}
\end{figure}

The selection as shown in Figure \ref{fig:3a} begins by collapsing the 690 000 raw modules down to only those that can actually synthesise and contribute meaningfully to timing analysis. First, every module is run through a standard RTL synthesis flow (e.g. Yosys or Synopsys DC) to weed out non-RTL constructs, unsupported system tasks, or modules that fail elaboration. Next, we remove near-duplicates by canonicalizing whitespace or AST structure and discarding any pair with more than 95 \% token overlap. The surviving pool is then annotated via a lightweight synthesis pass at a fixed clock constraint. For each module we record total gate count, flip-flop count, estimated maximum combinational depth, rough path-count, and preliminary TNS/WNS. These metrics let us define four orthogonal “stratification axes”: structural complexity (small to large gate counts), statefulness (pure combinational vs. flip-flop-heavy), timing difficulty (modules that already meet timing vs. those that violate), and functional domain (ALUs, FIFOs, FSMs, memories, arbiters, etc.). By explicitly marking where each module lies along these axes, we ensure that later sampling can enforce coverage across all dimensions of interest.

With this enriched metadata in hand, we turn to embedding and clustering to capture code-style diversity before final selection. Each module is embedded via a $\ell_{2}$-normalized bag-of-gates fingerprint. These vectors are then clustered (e.g. K-means) into roughly 100 buckets so that similar naming conventions, structural patterns, or design idioms end up grouped together. From each cluster, we perform stratified sampling so we divide modules into bins by gate-count ranges (0–.2 k, 0.2–0.5 k, 0.5–1 k, 1 k+), into slack-difficulty tiers (easy vs. hard), and by domain labels, then draw proportional samples from each bin. We also deliberately over-sample rare but important categories (like bus arbiters or CRC engines) to ensure the model encounters sufficient examples of edge-case architectures. The result is a 60k-module corpus that balances overall size with rich, multi-dimensional coverage, giving our two-stage LLM the varied, representative training set it needs to generalize robustly to unseen Verilog designs.

\subsection{Proposed model architecture}
The WNS is the most critical single-path violation; the TNS is the sum of all path violations.  For a combinational path \(i\) the STA report gives the required time \(R_i\) and the arrival time \(A_i\).  Their difference is the path slack
\begin{equation}
S_i = R_i - A_i .
\end{equation}
Only paths with \(S_i < 0\) contribute to the global metrics
\begin{equation}
\label{eq:tnswns}
\mathrm{TNS} = \sum_{i\,:\,S_i<0} S_i ,
\qquad
\mathrm{WNS} = \min_{i} S_i .
\end{equation}

\begin{figure*}[]
    \centering
    \includegraphics[width=0.8\linewidth]{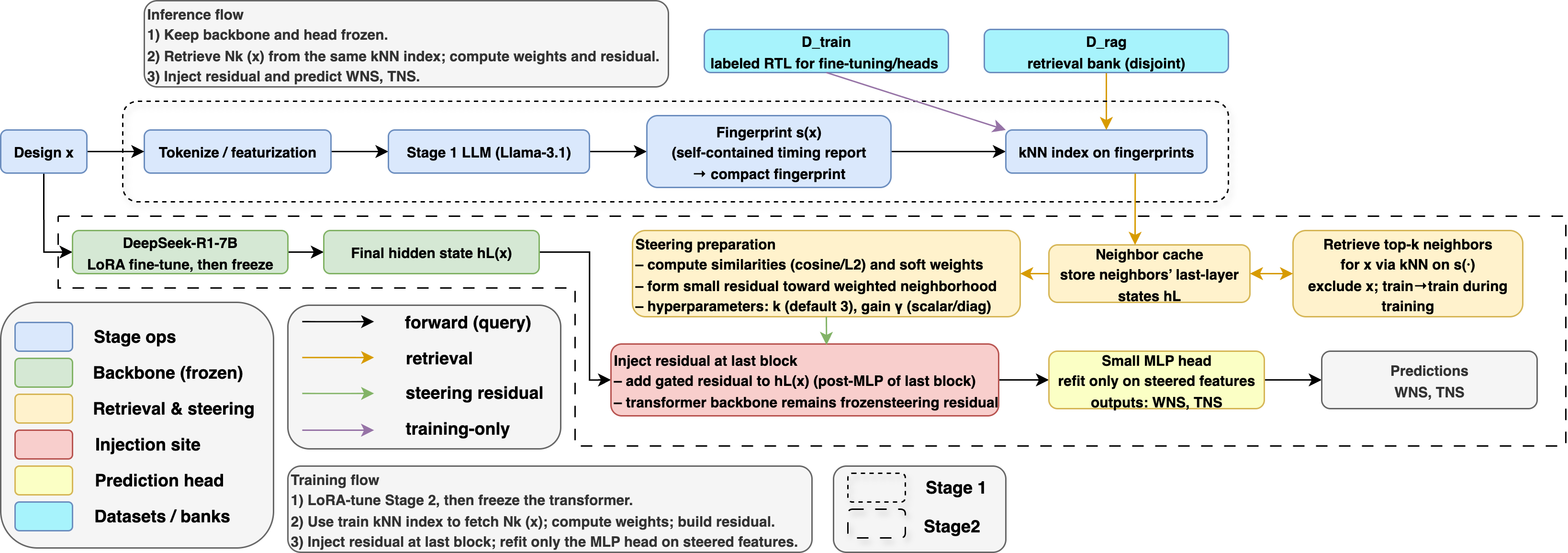}
    \captionsetup{justification=centering}
    \caption{Two-stage TimingLLM: fingerprint generation and retrieval-steered timing prediction.}
    \label{fig:2a}
\end{figure*}

TimingLLM predicts \(\mathrm{TNS}\) and \(\mathrm{WNS}\) directly from Verilog code using two consecutive LLM stages, providing timing estimation before full synthesis. The pipeline is illustrated in Figure \ref{fig:2a}. The neural models in the stages are trained on a labeled training set and use a separate retrieval set at query time. The datasets are disjoint by construction. \(\mathcal{D}_{\text{train}}=\{(x_j, y_j)\}_{j=1}^{N}\) for RTL and labels \(y_j=(\mathrm{TNS}_j,\mathrm{WNS}_j)\) used for fine-tuning, and \(\mathcal{D}_{\text{rag}}=\{x^{\mathrm{rag}}_m\}_{m=1}^{M}\) for a distinct corpus used only for similarity search; \(\mathcal{D}_{\text{rag}}\) is never used to update model weights.

Stage 1 is a fine-tuning model of the Llama 3.1 backbone on a large collection of Verilog modules whose synthesis reports are known.
The expected output of the LLM in this stage is not the final WNS or TNS numbers; instead, the model is fine-tuned to reproduce the entire timing reasoning, line by line, for every critical path. During training, the prompt contains the RTL source, and the target text is the corresponding STA report. Consequently, the model learns the semantics of \eqref{eq:tnswns} and can regenerate a synthetic, human-readable chain of thought that lists each critical path, its local slack, and the intermediate node delays. Given \(x\in\mathcal{D}_{\text{train}}\) and its tool report \(\hat r\), the LLM is fine-tuned to maximize the conditional likelihood of \(r\) given \(x\). At use time, Stage 1 receives a query RTL \(x\) and produces a synthetic timing report \(\hat r(x)\). This report is converted into a compact, timing-aware fingerprint by extracting statistics that summarize the report’s analysis context, endpoint and path types, critical-path topology and arc patterns, delay composition, parasitic/RC hints, and clock effects. Let \(\phi(\cdot)\in\mathbb{R}^{d}\) denote this feature extractor applied to \(\hat r(x)\). The fingerprint is \(\ell_2\)-normalized to make inner products meaningful:

\vspace{-2pt}
\begin{equation}
\label{eq:fingerprint}
\mathbf{s}(x) \;=\; \frac{\phi\bigl(\hat r(x)\bigr)}{\bigl\lVert \phi\bigl(\hat r(x)\bigr)\bigr\rVert_2}
\;\in\;\mathbb{R}^{d}.
\end{equation}
\vspace{-2pt}

\begin{figure}[h]
    \centering
    \includegraphics[width=0.55\linewidth]{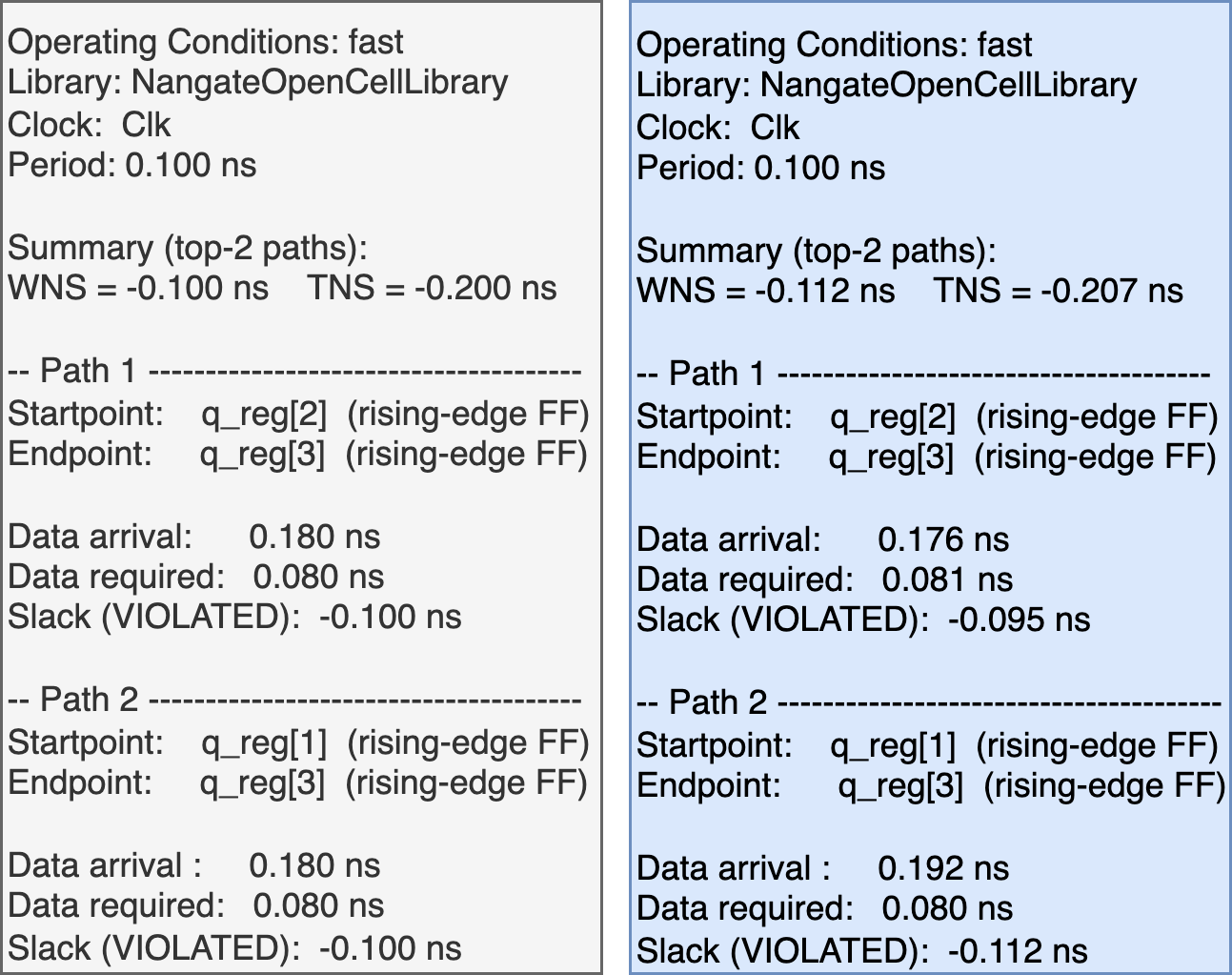}
    \captionsetup{justification=centering}
    \caption{Top-2 setup paths: STA vs. Stage-1 LLM agreement within 12 ps (WNS), 7 ps (TNS).}
    \label{fig:4a}
\end{figure}

Intuitively, \(\mathbf{s}(x)\) says what kind of timing problem this RTL looks like without requiring tool runs at query time. Figure \ref{fig:4a} presents a simplified view of the Stage~1 output, which is a list of paths with their local slack values only.
It omits gate identities, cell libraries, and full arrival/required\mbox{-}time breakdowns that appear in the actual report. The purpose is to show, in isolation, the core signal extracted in Stage~1, enumerated paths and their local slack, and to make clear how this signal feeds the normalized fingerprint that drives retrieval and steering in Stage~2. Fingerprints are built upon the report.

Stage 2 uses an LLM and treats the task as real-valued regression. In this paper, without loss of generality, we utilize DeepSeek-R1-7B model for the second stage. 
This stage performs regression to \(\mathrm{TNS}\) and \(\mathrm{WNS}\) and incorporates retrieval and steering. First, a baseline is fit on \(\mathcal{D}_{\text{train}}\) so the LLM produces a last-layer representation (logits) \(h_L(x;\Theta)\in\mathbb{R}^{d_h}\) which is mapped by a small MLP head \(g_\psi\) to two numbers (i.e., TNS and WNS). The baseline loss on \(\mathcal{D}_{\text{train}}\) is the mean squared error on the two outputs. To add retrieval, the synthetic-report fingerprints \(\mathbf{s}(\cdot)\) are precomputed for all items in \(\mathcal{D}_{\text{rag}}\). Given a query \(x\), we compute \(\mathbf{s}(x)\) by \eqref{eq:fingerprint} and retrieve the top-\(k\) neighbors \(\mathcal{N}_k(x)\subset\mathcal{D}_{\text{rag}}\) by inner product similarity
\begin{equation}
\label{eq:sim}
\mathrm{sim}\!\left(x, x^{\mathrm{rag}}\right) \;=\; \mathbf{s}(x)^{\!\top}\,\mathbf{s}\!\left(x^{\mathrm{rag}}\right).
\end{equation}
These neighbors provide structurally similar code samples whose internal LLM activations will steer the query’s representation. We convert similarities into nonnegative weights:
\begin{equation}
\label{eq:weights}
w_i \;=\; \frac{\exp\!\bigl(\mathrm{sim}(x,x_i^{\mathrm{rag}})\bigr)}{\sum_{x_j^{\mathrm{rag}}\in\mathcal{N}_k(x)} \exp\!\bigl(\mathrm{sim}(x,x_j^{\mathrm{rag}})\bigr)} \quad \text{for } x_i^{\mathrm{rag}}\in\mathcal{N}_k(x).
\end{equation}

\begin{figure}[h]
    \centering
    \includegraphics[width=0.99\linewidth]{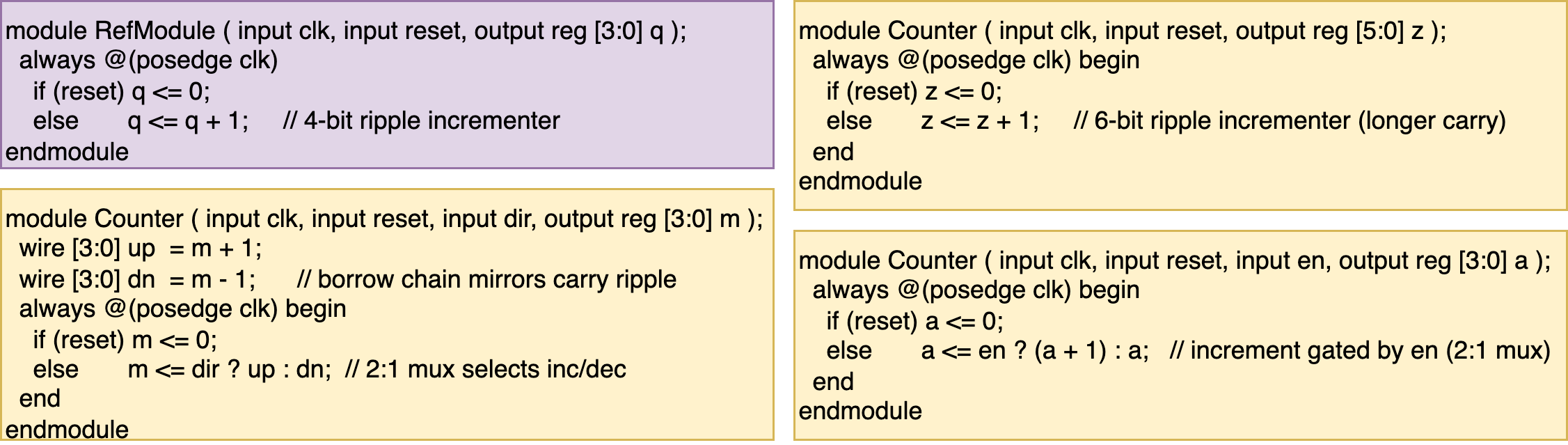}
    \captionsetup{justification=centering}
    \caption{Qualitative retrieval for feature-space steering}
    \label{fig:1a}
\end{figure}

Figure \ref{fig:1a} offers a qualitative snapshot of the retrieval stage. Given a query RTL module (purple box), our $\ell_{2}$-normalized structural--timing fingerprint retrieves the top-$k$ neighbors from the available designs in the retrieval bank (yellow boxes); these neighbors exhibit shared structural idioms and similar critical-path characteristics, indicating that proximity in fingerprint space aligns with timing-relevant similarity. 


Next, we forward each neighbor \(x_i^{\mathrm{rag}}\) through the LLM to read its last-layer activation \(h_L(x_i^{\mathrm{rag}};\Theta)\). Let \(h=h_L(x;\Theta)\) denote the query’s activation. We form a neighbor residual that points from \(h\) toward the weighted neighborhood:
\begin{equation}
\label{eq:residual}
v \;=\; \sum_{x_i^{\mathrm{rag}}\in\mathcal{N}_k(x)} w_i\,\Bigl(h_L(x_i^{\mathrm{rag}};\Theta) - h\Bigr).
\end{equation}
The steered representation adds a small, learned fraction of this residual to \(h\). Two equivalent parameterizations are common, which can be a single scalar gain \(\gamma_0\) shared across features, or a diagonal feature-wise gate \(\gamma\in\mathbb{R}^{d_h}\). The general form is
\begin{equation}
\label{eq:steer}
h' \;=\; h \;+\; \gamma \odot v,
\end{equation}
where \(\odot\) denotes elementwise multiplication (for a scalar \(\gamma_0\) this reduces to \(h' = h + \gamma_0 v\)). The purpose of \eqref{eq:steer} is to bias the query’s representation toward regions of representation space that historically correspond to similar timing patterns. The two-stage architecture, therefore, splits the task into an interpretable reasoning component and a compact regression component. Stage 1 learns how timing analysis works and explains its conclusions, while Stage 2 distils explanations, plus structural priors from related designs, into the two headline numbers required by place-and-route closure. Because retrieval selects neighbors from fingerprint similarity, no proprietary numeric labels leak across datasets, making the approach portable to new technology libraries or design styles.

\subsection{Training process}
We apply low-rank adapters (LoRA) to all attention blocks of the first-stage LLM and fine-tune the model to learn to generate the required information.
We have followed a similar LoRA approach for the LLM of the second stage. However, due to the existence of the steering path, its training is more complex compared to that of Stage 1. First, we LoRA fine-tune the Stage~2 LLM together with the regression head and save the resulting backbone parameters $\Theta$.
So to avoid any training–serving mismatch, the LLM parameters \(\Theta\) and the steering hyperparameter (\(\gamma\)) are then frozen. We recompute \(h'_j\) for every \((x_j,y_j)\in\mathcal{D}_{\text{train}}\) using the same retrieval and steering procedure, and we refit only the MLP head on these steered features so the head learns exactly the input distribution it will see at test time. Denoting the head by \(g_\psi:\mathbb{R}^{d_h}\to\mathbb{R}^{2}\) and the two predicted outputs by \(\widehat{\mathrm{TNS}},\widehat{\mathrm{WNS}}\), the head refit minimizes

\begin{equation}
\label{eq:head}
\frac{1}{N}\sum_{j=1}^{N}
\left\|
\begin{bmatrix}
\mathrm{TNS}_j \\[2pt] \mathrm{WNS}_j
\end{bmatrix}
-
g_\psi\!\bigl(h'_j\bigr)
\right\|_2^{2}.
\end{equation}

This step is lightweight, but crucial because adding \eqref{eq:steer} changes the feature distribution, and refitting \(g_\psi\) eliminates that distribution shift. At test time the flow exactly mirrors training so given an unseen RTL \(x_\star\), Stage 1 generates \(\hat r(x_\star)\) and the fingerprint \(\mathbf{s}(x_\star)\) by \eqref{eq:fingerprint}. Then nearest neighbors \(\mathcal{N}_k(x_\star)\subset\mathcal{D}_{\text{rag}}\) and weights \(w_i\) are computed using \eqref{eq:sim} and \eqref{eq:weights}; activations of \(x_\star\) and its neighbors are combined to produce \(h'_\star\) via \eqref{eq:residual}–\eqref{eq:steer}; finally the frozen head \(g_\psi\) outputs \(\bigl(\widehat{\mathrm{TNS}},\widehat{\mathrm{WNS}}\bigr)=g_\psi(h'_\star)\). Because \(\mathcal{D}_{\text{rag}}\) is disjoint from \(\mathcal{D}_{\text{train}}\) and similarity is computed only from Stage-1 report fingerprints, retrieval supplies structural priors without leaking training labels, while steering provides a controlled way to align the query representation with patterns observed in similar modules. Across all phases, the only trainable parameters are the LoRA adapters and the small regression head. Stage~1 updates adapters only. Stage~2 updates adapters and head initially, then head alone on steered features. The base LLM weights remain frozen at all times, which reduces the memory footprint. Items in $\mathcal{D}_{\text{rag}}$ are never used to compute a loss or update weights. During the initial Stage~2 fit and the head refit on steered features, the loss is computed only on $\mathcal{D}_{\text{train}}$ so $\mathcal{D}_{\text{rag}}$ is accessed by forward pass to obtain neighbor activations $h_i$ (labels are not read), and gradients are not taken with respect to any $\mathcal{D}_{\text{rag}}$ example.

\section{Results}

\subsection{Dataset Characterization}

\begin{figure}[h]
    \centering
    \captionsetup{justification=centering}
    \begin{subfigure}[t]{0.48\linewidth}
        \centering
        \includegraphics[width=\linewidth]{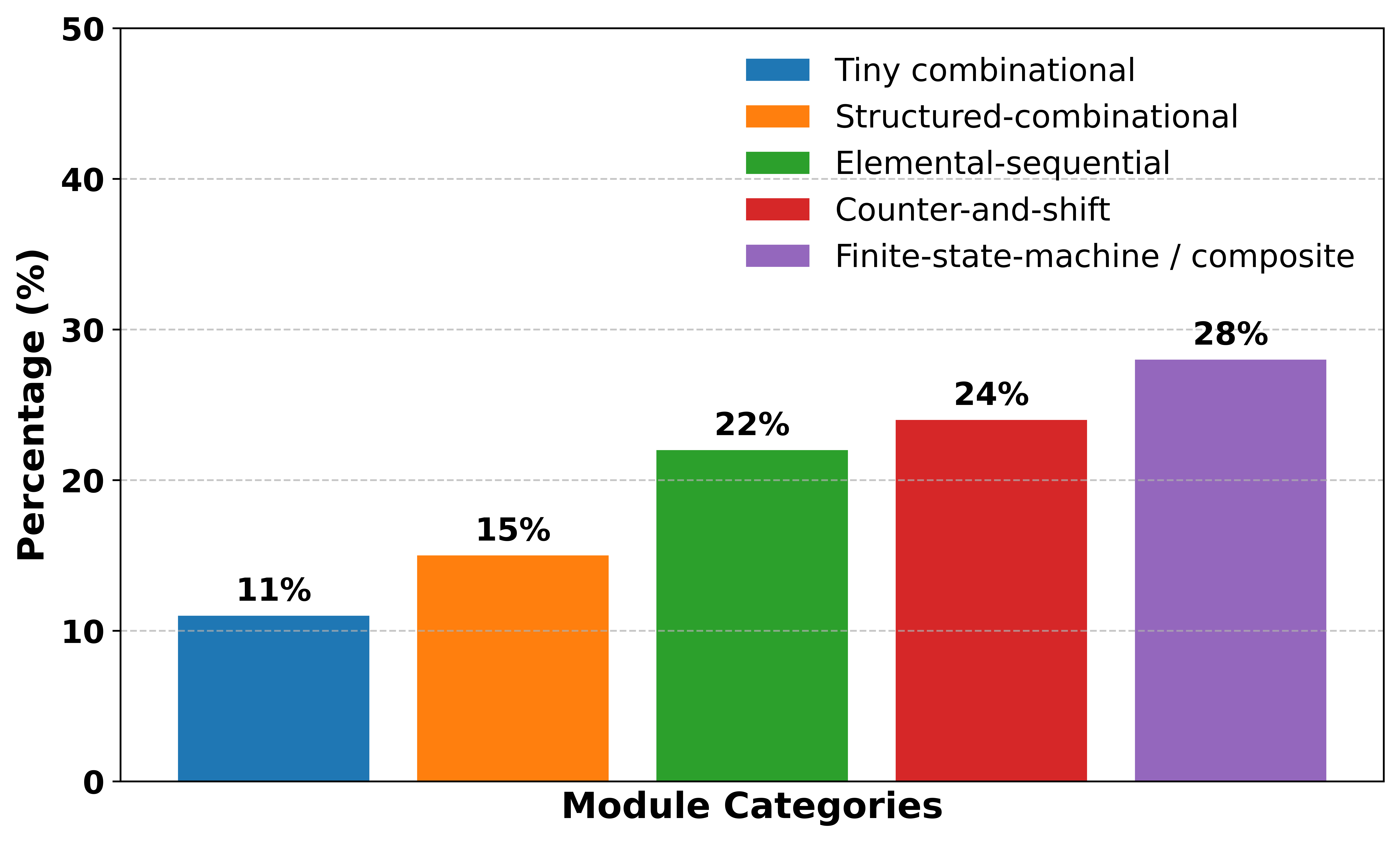}
        \caption{}
        \label{fig:5}
    \end{subfigure}\hspace{0.01\linewidth}
    \begin{subfigure}[t]{0.48\linewidth}
        \centering
        \includegraphics[width=\linewidth]{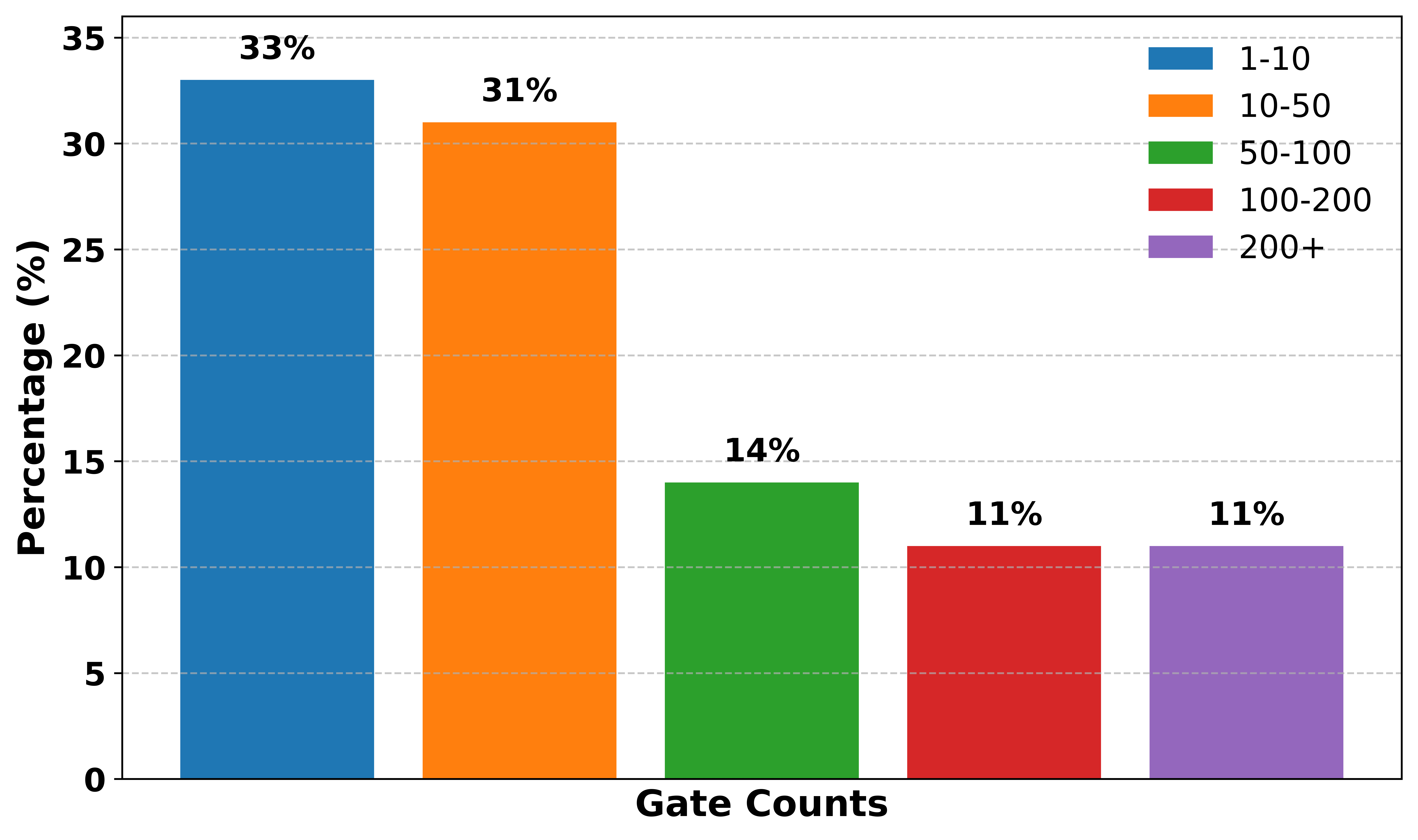}
        \caption{}
        \label{fig:6}
    \end{subfigure}
    \caption{The details of employed dataset}
    \label{fig:arch_flow}
\end{figure}

We present a 60k-module corpus organized into five functional tiers Figure \ref{fig:5}. Tiny-combinational modules (11\%) provide basic gate-level wiring exercises, while Structured-combinational modules (15\%) cover multiplexers, adders and comparators. Elemental-sequential modules (22\%) involve single-stage storage primitives, and Counter-and-shift modules (24\%) encompass various counting and shifting mechanisms. The largest tier, which is Finite-state-machine/composite modules (28\%), integrates control and datapath blocks. This distribution reflects a deliberate progression from elementary combinational logic toward increasingly stateful designs, ensuring broad coverage of timing-critical behavior and foundational constructs.

The distribution of synthesized gate counts across 60k-modules corpus exhibits a pronounced concentration in small-scale designs so 33\% of modules contain between 1 and 10 gates, and 31\% span 10–50 gates, together accounting for two-thirds of the dataset. Modules with moderate complexity (50–100 gates) comprise 14\%, while designs with 100–200 gates and 200+ gates each represent 11\%. Figure \ref{fig:6} summarizes it. This graded allocation of resources ensures comprehensive coverage from trivial combinational snippets through increasingly intricate sequential and control-datapath constructs, thereby enabling systematic evaluation of structural and timing representations as a function of module scale.

\subsection{Experimental Setup}
We use three disjoint sets: (i) an internal training set $\mathcal{D}_{\text{train}}$ of 40{,}000 modules to fit model weights; (ii) an internal retrieval-only memory $\mathcal{D}_{\text{rag}}$ of 20{,}000 modules used solely for steering; and (iii) an external test set $\mathcal{D}_{\text{test}}=$VerilogEval \cite{liu2023verilogeval} for all reported metrics. The sets are identity/near-duplicate disjoint, and VerilogEval is never used during training or retrieval bank construction. Retrieval keys are computed from the synthetic Stage~1 report (no ground-truth labels), and neighbors for both training-time steering and test-time steering are drawn exclusively from $\mathcal{D}_{\text{rag}}$.
We used $R$ metric which is the Pearson correlation coefficient between ground-truth slacks $y_i$ and predictions $\hat{y}_i$ and $\mathrm{MAPE}$ metric which is the mean absolute percentage error in percent to evaluate and compare our results.

\begin{align}
\mathrm{MAPE} &=
\frac{100}{N}\sum_{i=1}^{N}
\biggl|\frac{y_i-\hat{y}_i}{y_i}\biggr|
\end{align}

The first stage fine-tunes Llama-3.1-8B-Instruct \cite{llama3_2024} on fourty-thousand synthesised Verilog modules.  Training is parameter-efficient so only LoRA adapters with rank~16, scaling~$\alpha=16$, and dropout~0.05 are added to the four attention projections, while all backbone weights stay frozen. Each sample provides the raw RTL as input and the full static-timing report as target text, tokenized to at most 1024 tokens. Optimization uses AdamW with a learning rate of $5\!\times\!10^{-5}$, cosine decay, and an effective batch size of 32. The second stage fine-tunes DeepSeek-R1-7B \cite{deepseek_r1_2025} to regress the natural-logarithm of TNS and WNS. LoRA again modifies only the attention projections, this time with rank 16 and $\alpha=8$. Prompts are capped at 1024 tokens. The model trains for ten epochs on fourty-thousand labelled modules, with a twenty-thousand-design retrieval bank held in GPU memory. AdamW (\(10^{-4}\) learning rate, batch size 8, fp16) minimizes a $L_{2}$ loss; the checkpoint with the lowest error on a held-out set is kept, its adapters are merged into the backbone, and the combined model outputs $\widehat{\mathrm{TNS}}$ and $\widehat{\mathrm{WNS}}$ given only the RTL source. TimingLLM consumes no .lib/PDK features at inference. Stage~1 emits structure–timing fingerprints from RTL; Stage~2 predicts $(\mathrm{WNS},\mathrm{TNS})$ using retrieval whose keys are tagged with lightweight metadata tokens (node, lib, proc, vdd, temp, delay, and libtag). For the main VerilogEval results, all PyraNet training labels and VerilogEval test labels are obtained from Synopsys Design Compiler (compile\_ultra) targeting the Nangate45 Open Cell Library (CCS) \cite{nangate45} at the typical corner. In our cross-technology study, we consider three Liberty libraries—Nangate45, Nangate15 \cite{nangate15_isqed_2015}, and ASAP7 \cite{asap7_iccad_2017}—and, for each library, we evaluate two PVT corners, a typical corner and a worst-case slow setup (SS, low $V_{\mathrm{DD}}$, high temperature) corner. Across all of these settings we keep Stage~1 and the retrieval bank (built once from PyraNet synthesized to Nangate45 at the typical corner) fixed, and adapt only a two-layer regression head for each target (library, corner) pair by re-synthesizing 1000 PyraNet-train modules under that setting and refitting the head on those labels, while evaluating on VerilogEval synthesized under the same (library, corner).

\subsection{Discussion}

\begin{table}[htbp]
\centering
\renewcommand{\arraystretch}{0.8} 
\caption{WNS/TNS prediction (R↑, MAPE↓).}
\vspace{-8pt}
\begin{tabular}{l||cc|cc}
\hline
Method       & \multicolumn{2}{c|}{WNS}      & \multicolumn{2}{c}{TNS}      \\
\cline{2-5}
             & R       & MAPE    & R       & MAPE    \\
\hline
RTL‐Timer    & 0.87    & 19\%    & 0.92    & 28\%    \\
MasterRTL    & 0.86    & 20\%    & 0.91    & 30\%    \\
CircuitFusion& 0.89    & 14\%    & 0.95    & 18\%    \\
\textbf{TimingLLM}& \textbf{0.91} & \textbf{12\%} & \textbf{0.97} & \textbf{16\%} \\
\hline
\end{tabular}
\label{tab:wns-tns-comparison}
\end{table}

Table \ref{tab:wns-tns-comparison} shows that while both RTL-Timer and MasterRTL achieve respectable WNS/TNS correlations (R = 0.87–0.92) their percentage errors remain relatively high (MAPE = 19–30\%), reflecting modest calibration on timing‐slack prediction. CircuitFusion narrows that gap, its embeddings-based regressor yields stronger correlations (R = 0.89/0.95) and cuts MAPE nearly in half for WNS (14\%) and substantially for TNS (18\%). Our method outperforms all baselines, delivering the highest linear agreement (R=0.91 for WNS, 0.97 for TNS) alongside the lowest average percentage errors (12\% and 16\%, respectively), which demonstrates its superior accuracy and consistency in predicting worst‐case and TNS.

Tables~\ref{tab:wns-category}–\ref{tab:tns-category} show that TimingLLM achieves the best R and lowest MAPE across all five VerilogEval categories for both WNS and TNS. The largest margins over the next-best (CircuitFusion) appear on \emph{FSM/composite} and \emph{elemental-sequential} blocks, while gaps are smallest on \emph{tiny/structured combinational}. CircuitFusion weakens on control-heavy designs; RTL-Timer degrades on FSM; MasterRTL trails consistently. These trends indicate the robustness of our method to all categories.

\begin{table}[htbp]
\centering
\renewcommand{\arraystretch}{0.8} 
\scriptsize
\setlength{\tabcolsep}{3pt}\renewcommand{\arraystretch}{0.6}
\caption{Per-category WNS: correlation (R) / MAPE.}
\vspace{-8pt}
\resizebox{\columnwidth}{!}{%
\begin{tabular}{lcccc}
\toprule
Category & \textbf{TimingLLM} & CircuitFusion & RTL-Timer & MasterRTL \\
\midrule
Tiny comb.                 & \textbf{0.90 / 11\%} & 0.89 / 12\% & 0.87 / 16\% & 0.85 / 18\% \\
Struct. comb.              & \textbf{0.92 / 12\%} & 0.90 / 14\% & 0.88 / 19\% & 0.86 / 20\% \\
Elem.-seq.                 & \textbf{0.91 / 13\%} & 0.88 / 16\% & 0.86 / 21\% & 0.84 / 23\% \\
Count./shift               & \textbf{0.93 / 11\%} & 0.90 / 13\% & 0.88 / 18\% & 0.86 / 20\% \\
FSM / comp.                & \textbf{0.90 / 14\%} & 0.86 / 18\% & 0.85 / 22\% & 0.83 / 24\% \\
\bottomrule
\end{tabular}}
\label{tab:wns-category}
\end{table}

\begin{table}[htbp]
\centering
\scriptsize
\setlength{\tabcolsep}{3pt}\renewcommand{\arraystretch}{0.6}
\caption{Per-category TNS: correlation (R) / MAPE.}
\resizebox{\columnwidth}{!}{%
\begin{tabular}{lcccc}
\toprule
Category & \textbf{TimingLLM} & CircuitFusion & RTL-Timer & MasterRTL \\
\midrule
Tiny comb.                 & \textbf{0.97 / 15\%} & 0.95 / 17\% & 0.92 / 26\% & 0.91 / 28\% \\
Struct. comb.              & \textbf{0.97 / 16\%} & 0.95 / 18\% & 0.93 / 28\% & 0.92 / 30\% \\
Elem.-seq.                 & \textbf{0.96 / 17\%} & 0.94 / 19\% & 0.92 / 31\% & 0.91 / 33\% \\
Count./shift               & \textbf{0.98 / 15\%} & 0.95 / 17\% & 0.93 / 27\% & 0.92 / 29\% \\
FSM / comp.                & \textbf{0.96 / 18\%} & 0.93 / 21\% & 0.91 / 33\% & 0.90 / 35\% \\
\bottomrule
\end{tabular}}
\label{tab:tns-category}
\end{table}

\subsection{Memory and Runtime}


TimingLLM pipeline completes in about 323 seconds end-to-end: 0.55 seconds preprocessing, 317 seconds for Stage 1, and 5.5 seconds for Stage 2, with GPU memory 17–18 GB for Stage 1 and 14.5–15.5 GB for Stage 2. For MasterRTL, end-to-end runtime is about 514 seconds, dominated by SOG construction and feature extraction. RTL-Timer reports total modeling cost of about 4\% of default synthesis (3.2\% from AIG construction), yielding about 411 seconds over VerilogEval. CircuitFusion requires about ten hours pre-training on four A4000 GPUs and five minutes to fine-tune per task on a server with 512 GB RAM, leading to 467 seconds over this dataset. To ensure fairness, we ran authors' implementations on our workstation under uniform setup (RTX A6000 48 GB; 16-core CPU 3.5 GHz; 128 GB RAM). The results show our superiority in speed while achieving better R and lower MAPE across VerilogEval.

\subsection{Cross-Library and Cross-Corner Adaptation}

To assess portability across technology libraries and PVT corners, we evaluate three libraries and two PVT corners per library, as summarized in Table~\ref{tab:cross_lib_corner}. For each target (library, corner) pair $(L',c')$ we perform a head-only adaptation where we re-synthesize 1000 PyraNet-train modules under $(L',c')$ with Design Compiler and refit the two-layer regression head on these labels while keeping the Stage~1 LLM, the structural fingerprint, and the Nangate45-based retrieval bank fixed. The 1000 modules are selected once using a simple stratified sampling over design size and depth, without using any timing labels, and the same indices are reused for all libraries and corners. The test set is VerilogEval synthesized under the same $(L',c')$; we report correlation $R$ and MAPE for WNS and TNS. This protocol uses roughly $2\%$ of the PyraNet training set per setting and does not require rebuilding the retrieval bank or retraining the timing reasoner.

Table~\ref{tab:cross_lib_corner} shows that TimingLLM maintains roughly the same accuracy across all three libraries and both PVT corners. In other words, changing the library or moving from typical to a slow corner does not materially degrade performance. Additionally, across all libraries and corners, the adapted TimingLLM remains better than per-library RTL-Timer, CircuitFusion, and MasterRTL models.

\begin{table}[h]
\centering
\renewcommand{\arraystretch}{0.5} 
\caption{Cross-library/corner WNS/TNS prediction (R↑, MAPE↓).}
\vspace{-8pt}
\begin{tabular}{l l||cc|cc}
\hline
Lib & Corner & \multicolumn{2}{c|}{WNS} & \multicolumn{2}{c}{TNS} \\
    &        & R & MAPE & R & MAPE \\
\hline
Nangate45 & typ & 0.91 & 12\% & 0.97 & 16\% \\
Nangate45 & slow & 0.90 & 13\% & 0.96 & 16\% \\
Nangate15 & typ & 0.90 & 12\% & 0.96 & 17\% \\
Nangate15 & slow & 0.90 & 13\% & 0.96 & 17\% \\
ASAP7     & typ & 0.91 & 12\% & 0.97 & 17\% \\
ASAP7     & slow & 0.90 & 12\% & 0.95 & 16\% \\
\hline
\end{tabular}
\label{tab:cross_lib_corner}
\end{table}
\subsection{Ablation Study}

The ablation in Table \ref{tab:stage_ablation_compact} confirms that the two stages are complementary. When we train only Stage 1, fine-tuning the Llama timing reasoner without the retrieval-augmented regression, the correlation coefficient~\(R\) for both WNS and TNS drops and MAPE rises, indicating that path-level explanations alone do not translate directly into precise numeric slack estimates. Conversely, bypassing the chain-of-thought and relying solely on Stage 2 regression also degrades performance, because the model loses the positive effect of RAG modules and their steering values. The full pipeline, which fuses Stage 1’s explanatory reports with Stage 2’s structurally guided regression, recovers the highest correlation and the lowest MAPE, demonstrating that accurate slack prediction emerges only when both reasoning and retrieval signals are present. The results in Table \ref{tab:k} demonstrate a clear improvement in both correlation coefficient \(R\) and mean absolute percentage error (MAPE) as the number of retrieved neighbours increases from \(k=1\) to \(k=3\).  The optimal performance at \(k=3\) indicates that three similar, timing-labelled modules provide the most effective context for the regression model, yielding the highest predictive accuracy for both TNS and WNS. Beyond this point, increasing to \(k=5\) introduces less relevant exemplars, slightly degrading performance and suggesting diminishing returns.

\begin{table}[htbp]
\centering
\caption{Stage removal ablation study.}
\label{tab:stage_ablation_compact}
\setlength{\tabcolsep}{3pt}
\scriptsize
\resizebox{\columnwidth}{!}{
\begin{tabular}{lcccc}
\toprule
Method & $\boldsymbol{R_{\text{WNS}}\uparrow}$ & $\boldsymbol{\mathrm{MAPE}_{\text{WNS}}\downarrow}$ & $\boldsymbol{R_{\text{TNS}}\uparrow}$ & $\boldsymbol{\mathrm{MAPE}_{\text{TNS}}\downarrow}$ \\
\midrule
Stage 1 only (chain-of-thought reasoning) & 0.83 & 21\% & 0.85 & 19\% \\
Stage 2 only (Regression, no CoT)   & 0.76 & 25\% & 0.78 & 24\% \\
Full pipeline (Stage 1 + Stage 2)         & \textbf{0.91} & \textbf{12\%} & \textbf{0.97} & \textbf{16\%} \\
\bottomrule
\end{tabular}
}
\end{table}

\begin{table}[htbp]
\centering
\footnotesize
\setlength{\tabcolsep}{3pt}      
\renewcommand{\arraystretch}{0.5}
\caption{WNS/TNS performance vs.\ neighbour count $k$.}
\vspace{-8pt}
\label{tab:k}
\begin{tabular}{@{}ccccc@{}}
\toprule
\(k\) & \(R_{\mathrm{WNS}}\) & \(\mathrm{MAPE}_{\mathrm{WNS}}\) &
\(R_{\mathrm{TNS}}\) & \(\mathrm{MAPE}_{\mathrm{TNS}}\) \\
\midrule
1 & 0.85 & 19\% & 0.88 & 22\% \\
2 & 0.90 & 13\% & 0.95 & 17\% \\
\textbf{3} & \textbf{0.91} & \textbf{12\%} & \textbf{0.97} & \textbf{16\%} \\
5 & 0.90 & 14\% & 0.96 & 17\% \\
\bottomrule
\end{tabular}
\end{table}

As shown in Table~\ref{tab:gamma_abl_wns_tns}, when we steer at last layer and retrain the MLP head on the steered features, the feature-wise (diagonal) $\gamma\in\mathbb{R}^d$ achieves the best accuracy on both WNS and TNS. Among scalar settings, performance follows a shallow U-shape with an optimum around $\gamma{=}0.10$; larger magnitudes begin to hurt due to distribution shift. This supports using a $d$-dimensional gate when the head is retrained, while scalar $\gamma$ remains a stable, low-parameter baseline. 

\begin{table}[htbp]
\centering
\renewcommand{\arraystretch}{0.6} 
\caption{Steering gain $\gamma$ ablation: scalar vs. diagonal at last layer ($k{=}3$, head retrained).}
\vspace{-8pt}
\label{tab:gamma_abl_wns_tns}
\setlength{\tabcolsep}{3pt}
\scriptsize
\resizebox{\columnwidth}{!}{
\begin{tabular}{lcccccc}
\toprule
\textbf{Gamma type} & $\boldsymbol{\gamma}$ & $\boldsymbol{R_{\text{WNS}}\uparrow}$ & $\boldsymbol{\text{MAPE}_{\text{WNS}}\downarrow}$ & $\boldsymbol{R_{\text{TNS}}\uparrow}$ & $\boldsymbol{\text{MAPE}_{\text{TNS}}\downarrow}$ \\
\midrule
Scalar  & 0.02 & 0.88  & 14\%   & 0.94  & 18\% \\
Scalar  & 0.05 & 0.90  & 13\%   & 0.96  & 17\% \\
Scalar  & \textbf{0.10} & \textbf{0.905} & \textbf{12.5\%} & \textbf{0.965} & \textbf{16.7\%} \\
Scalar  & 0.20 & 0.900 & 12.8\% & 0.960 & 17.0\% \\
Scalar  & 0.30 & 0.892 & 13.5\% & 0.952 & 17.8\% \\
Scalar  & 0.40 & 0.880 & 14.5\% & 0.940 & 19.0\% \\
\midrule
Diagonal ($d$-dim) & learned & \textbf{0.91} & \textbf{12\%} & \textbf{0.97} & \textbf{16\%} \\
\bottomrule
\end{tabular}
}
\end{table}

Table~\ref{tab:layer_single_ablation} examines where to inject the diagonal d-dimensional steering vector $\gamma$ within the transformer. The retrieval configuration is fixed ($k=3$ cosine-weighted neighbors), the LLM is frozen, the residual is added post-MLP inside the chosen block, and the prediction head is refit on the steered features. Under these controls, steering at the last block $L_{28}$ gives the strongest accuracy-error trade-off for both objectives (highest $R_{\text{WNS}}$ and $R_{\text{TNS}}$ with the lowest $\text{MAPE}_{\text{WNS}}$ and $\text{MAPE}_{\text{TNS}}$). Moving the same steering budget to earlier blocks ($L_{27}$, $L_{24}$, $L_{14}$, $L_{7}$) yields a consistent degradation. In Table~\ref{tab:layer_double_ablation} We keep the overall strength fixed and use a split coefficient $\alpha\in[0,1]$ so an $\alpha$ share is applied at the later block in the pair $(L_b)$ and the remaining $(1-\alpha)$ share at the earlier block $(L_a)$, injected at the same post\mbox{-}MLP site as in the single\mbox{-}layer study.
Concentrating the entire strength on the last block (pair $L_{27}/L_{28}$ with $\alpha=1$) matches or slightly exceeds split configurations, while spreading it across blocks ($\alpha<1$) tends to dilute the effect.

\begin{table}[htbp]
\centering
\renewcommand{\arraystretch}{0.6} 
\caption{Single-layer steering across transformer blocks (diagonal $\gamma$, post-MLP, $k{=}3$).}
\label{tab:layer_single_ablation}
\setlength{\tabcolsep}{3pt}
\scriptsize
\resizebox{\columnwidth}{!}{
\begin{tabular}{lcccccc}
\toprule
\textbf{Layer (block)} & \textbf{Site} & $\boldsymbol{R_{\text{WNS}}\uparrow}$ & $\boldsymbol{\text{MAPE}_{\text{WNS}}\downarrow}$ & $\boldsymbol{R_{\text{TNS}}\uparrow}$ & $\boldsymbol{\text{MAPE}_{\text{TNS}}\downarrow}$ \\
\midrule
$L_{28}$ (last)    & post-MLP   & \textbf{0.910} & \textbf{12.0\%} & \textbf{0.970} & \textbf{16.0\%} \\
$L_{27}$           & post-MLP   & 0.902 & 12.6\% & 0.964 & 16.7\% \\
$L_{24}$           & post-MLP   & 0.900 & 12.9\% & 0.962 & 17.0\% \\
$L_{14}$ (middle)  & post-MLP   & 0.894 & 13.4\% & 0.958 & 17.6\% \\
$L_{7}$  (early)   & post-MLP   & 0.887 & 14.1\% & 0.952 & 18.3\% \\
\bottomrule
\end{tabular}
}
\end{table}

\begin{table}[htbp]
\centering
\renewcommand{\arraystretch}{0.6} 
\caption{Two-layer steering with budget split $\alpha$ (diagonal $\gamma$, post-MLP, $k{=}3$).}
\vspace{-8pt}
\label{tab:layer_double_ablation}
\setlength{\tabcolsep}{3pt}
\scriptsize
\resizebox{\columnwidth}{!}{
\begin{tabular}{lcccccc}
\toprule
\textbf{Layers} & \boldmath$\alpha$ (later layer) & $\boldsymbol{R_{\text{WNS}}\uparrow}$ & $\boldsymbol{\text{MAPE}_{\text{WNS}}\downarrow}$ & $\boldsymbol{R_{\text{TNS}}\uparrow}$ & $\boldsymbol{\text{MAPE}_{\text{TNS}}\downarrow}$ \\
\midrule
$L_{28}+L_{27}$ & 1.00 / 0.00 & \textbf{0.91} & \textbf{12\%} & \textbf{0.97} & \textbf{16\%} \\
$L_{28}+L_{27}$ & 0.75 / 0.25 & 0.909 & 12.1\% & 0.969 & 16.1\% \\
$L_{28}+L_{27}$ & 0.50 / 0.50 & 0.908 & 12.2\% & 0.969 & 16.2\% \\
$L_{28}+L_{24}$ & 0.75 / 0.25 & 0.909 & 12.1\% & 0.968 & 16.2\% \\
$L_{27}+L_{14}$ & 0.75 / 0.25 & 0.902 & 12.6\% & 0.965 & 16.6\% \\
\bottomrule
\end{tabular}
}
\end{table}

\section{Conclusion}

We presented TimingLLM, a two-stage, tool-free RTL timing predictor in which Stage~1 produces an STA-style report and an $\ell_2$-normalized structural–timing fingerprint to index a disjoint 20k-module retrieval bank, and Stage~2 regresses WNS/TNS with a lightweight neighbor-conditioned residual at the final transformer block. On VerilogEval it attains $R_{\text{WNS}}{=}0.91$ (MAPE 12\%) and $R_{\text{TNS}}{=}0.97$ (MAPE 16\%), completing end-to-end in $\sim$323,s on one RTX~A6000 versus $\sim$514,s (MasterRTL), $\sim$411,s (RTL-Timer), and $\sim$467,s (CircuitFusion) under a uniform setup. A cross-technology study shows that the same trained model can be adapted to three Liberty libraries and two PVT corners by refitting only a two-layer regression head on 1000 modules per setting, keeping Stage~1 and the retrieval bank fixed and maintaining essentially the same accuracy.

\bibliographystyle{unsrt}
\bibliography{myref}

\end{document}